\documentclass[journal]{IEEEtran}
\usepackage{amsfonts}
\newtheorem{remark}{Remark}
\usepackage{bm}
\usepackage{amsmath}
\usepackage{makecell}
\usepackage{amssymb}
\usepackage{array}
\usepackage{cases}
\usepackage{algorithm}
\usepackage{algorithmic}
\usepackage{enumerate}
\usepackage{color,xcolor}
\usepackage{graphicx}
\usepackage{multirow,tabularx}
\usepackage{subfigure}
\usepackage{multirow}
\usepackage{hyperref}
\usepackage[compress]{cite}
\usepackage{setspace}

\makeatletter

\newcommand{\Rmnum}[1]{\expandafter\@slowromancap\romannumeral #1@}
\usepackage[vlined, ruled, algo2e]{algorithm2e}
\makeatother
 0

\begin{document}

\title{Variational Bayesian Inference Clustering Based Joint User Activity and Data Detection for Grant-Free Random Access in mMTC}

\author{Zhaoji Zhang,~\IEEEmembership{Member,~IEEE,} Qinghua Guo,~\IEEEmembership{Senior~Member,~IEEE,} Ying Li,~\IEEEmembership{Member,~IEEE,}\\ Ming Jin,~\IEEEmembership{Member,~IEEE,} and Chongwen Huang,~\IEEEmembership{Member,~IEEE}

\thanks{Zhaoji Zhang and Ying Li are with the School of Telecommunications Engineering, Xidian University, Xi'an 710071, China (email: zhaojizhang@xidian.edu.cn; yli@mail.xidian.edu.cn).
	
Qinghua Guo is with the School of Electrical, Computer and
Telecommunications Engineering, University of Wollongong, Wollongong,
NSW 2522, Australia (e-mail: qguo@uow.edu.au)

Ming Jin is with the Faculty of Electrical Engineering and Computer Science, Ningbo University, Ningbo 315211, China (e-mail: jinming@nbu.edu.cn).

Chongwen Huang is with the College of Information Science and Electronic Engineering, Zhejiang University, Hangzhou 310027, China, also with the International Joint Innovation Center, Zhejiang University, Haining 314400, China, and also with the Zhejiang Provincial Key Laboratory of Information Processing, Communication and Networking (IPCAN), Hangzhou 310027, China (e-mail: chongwenhuang@zju.edu.cn).

\emph{(Corresponding Author: Ying Li)}}}
\maketitle{
\begin{abstract}
T{\color{blue}ailor-made for massive connectivity and sporadic access, grant-free random access has become a promising candidate access protocol  for massive machine-type communications (mMTC). Compared with conventional grant-based protocols, grant-free random access skips the exchange of scheduling information to reduce the signaling overhead, and facilitates sharing of access resources to enhance access efficiency. However, some challenges remain to be addressed in the receiver design, such as unknown identity of active users and multi-user interference (MUI) on shared access resources. In this work, we deal with the problem of joint user activity and data detection for grant-free random access. Specifically, the approximate message passing (AMP) algorithm is first employed to mitigate MUI and decouple the signals of different users. Then, we extend the data symbol alphabet to incorporate the null symbols from inactive users. In this way, the joint user activity and data detection problem is formulated as a clustering problem under the Gaussian mixture model. Furthermore, in conjunction with the AMP algorithm, a variational Bayesian inference based clustering  (VBIC) algorithm is developed to solve this clustering problem. Simulation results show that, compared with  state-of-art solutions, the proposed AMP-combined VBIC (AMP-VBIC) algorithm achieves a significant performance gain in detection accuracy.}
\end{abstract}
\begin{IEEEkeywords}
Massive machine-type communications, grant-free, joint user activity and data detection, variational Bayesian inference, clustering, approximate message passing.
\end{IEEEkeywords}
\IEEEpeerreviewmaketitle
\section{Introduction}


\IEEEPARstart{I}nternet of Things (IoT) facilitates information exchange among objects in the physical world, and motivates the development for a diversity of novel applications, such as the smart city, smart grid, factory automation, etc. As an important constituent scenario in 5G, massive machine-type communications (mMTC)  has been proposed to accommodate diversified IoT services \cite{mmtc}. Compared with conventional scenarios, mMTC is characterized by (i) the massiveness and low activation probability of user equipments (UEs), (ii) short data packets from activated UEs, and (iii) demand for low power consumption by low-cost UEs. Furthermore, these features will become more prominent with the evolution of B5G and 6G. 
	
In medium access control (MAC) protocols, the random access mechanism configures connection setup for uplink transmission, i.e., the random access procedure allocates transmission resources to randomly activated UEs. However, the massiveness of UEs and shortage of uplink resources in mMTC have made random access a bottleneck problem for MAC designs \cite{RA1,RA2}. Existing random access schemes can be roughly divided into two categories, i.e. \emph{grant-based} and \emph{grant-free} schemes. In grant-based random access schemes, a handshaking procedure is needed to exchange the control signaling between the base station (BS) and active UEs. However, this handshaking procedure may incur prohibitively high signaling overhead for mMTC, which undermines the transmission efficiency of the small-sized data packets.

As an alternative to grant-based schemes, \emph{grant-free} random access has emerged in recent years. In grant-free schemes, the handshaking procedure is skipped, while active UEs can share the uplink access resources, and directly transmit their data packets without the grant from the BS. To ensure successful data recovery under grant-free random access, several critical problems need to be addressed at the BS. For example, the BS needs to solve the user-activity detection (UAD) problem to identify the active UEs, as well as the channel estimation (CE) problem to obtain the channel state information (CSI) for these active UEs. After that, the BS needs to solve the multi-user detection (MUD) problem to detect the data from active UEs. Considering different enabling techniques for grant-free random access, the state-of-art solutions to above-mentioned problems are reviewed as follows.

{\color{blue} \subsection{Grant-Free Random Access Enabled by MIMO and OFDM}	
As important enabling techniques for mMTC, the multiple input multiple output (MIMO) technique and orthogonal frequency division multiplexing (OFDM) technique can exploit spatial diversity and frequency diversity respectively to support the massive connectivity. On the other hand, the mobile traffic report \cite{report} shows that only a small fraction of UEs will be activated in typical IoT applications. To exploit this sparseness of active UEs, the framework of compressed sensing (CS)\cite{CS1,CS2} has received extensive studies. Under this CS framework, each UE is allocated with a unique pilot sequence, which will be transmitted with its data packet if this UE is activated. In this way, MIMO-enabled and OFDM-enabled grant-free random access share similar formulation of the detection problem, and the entire detection procedure is typically divided into two steps. Firstly, the \emph{joint UAD and CE} problem is formulated as a \emph{sparse-signal recovery problem}. Different CS algorithms have been proposed for this problem, such as the modified Bayesian CS algorithm \cite{mBayesian}, the block orthogonal matching pursuit (BOMP) algorithm \cite{BOMP}, the approximate message passing (AMP) algorithm \cite{YuweiAMP1,YuweiAMP2,YuweiAMP3}, the deep neural network-aided sparse Bayesian learning algorithm \cite{DNNSBL}. In the second step, the MUD problem can be readily addressed according to the UAD and CE results. 

\subsection{Grant-Free Random Access Enabled by Spreading}
The spreading technique serves as another enabling technique for mMTC with intriguing implementation feasibility. In spreading-enabled grant-free access mechanisms \cite{uniform1,AMPUADMUD,SISD,BSASP,JUICESD}, each data symbol is spread with a UE-specific spreading sequence, while all the spread symbols of each UE experience the same \emph{scalar} channel gain during transmission. In this way, the CE problem is significantly simplified, and spreading-enabled grant-free random access enjoys a much simpler problem formulation for receiver design. Then, different solutions have been proposed for the joint UAD and MUD problem. For example, an iterative order recursive least square (IORLS) algorithm \cite{uniform1} was proposed to exploit the joint sparsity of the data matrix to improve the detection accuracy. A joint expectation maximization and AMP (EM-AMP) algorithm was proposed in \cite{AMPUADMUD}, where the data matrix is detected from the received signal by the AMP algorithm \cite{AMP}, while the activity detection is addressed by the EM algorithm \cite{EM}. In addition, a structured iterative support detection (SISD) algorithm is proposed in \cite{SISD}. In \cite{BSASP}, a block sparsity adaptive subspace pursuit (BSASP) algorithm is proposed for the joint UAD and MUD problem, while the CE problem is addressed with a reference symbol. Recently, a joint UAD, CE, and signal detection (JUICESD) algorithm was proposed in \cite{JUICESD}, where the AMP algorithm is employed for signal detection and the detected signals are also used to refine the CE result.

These above-mentioned solutions \cite{uniform1,AMPUADMUD,SISD,BSASP,JUICESD} involve some infeasible assumptions or deficiencies. For example, the sparsity level, i.e. the \emph{exact} number of active UEs is assumed known to the BS in \cite{uniform1}, while the schemes in \cite{AMPUADMUD,SISD} require perfect knowledge on CSI at receiver (CSIR) even for inactive UEs. Such information is commonly unavailable in mMTC scenarios due to the massiveness and random activity of UEs. In addition, the subspace pursuit principle in \cite{BSASP} fails to address the inherent modulation constraint of data symbols, which undermines the data-detection accuracy. The UAD in \cite{JUICESD} relies on a non-deterministic detection threshold, while fine-tuning this threshold may incur tedious work under complicated mMTC scenarios. Recently, some advances on MUD techniques have inspired new ideas to tackle these deficiencies, and the details are explained in the next subsection. 

\subsection{Clustering and Variational Bayesian Inference for MUD}
It is noted that modulated data symbols are discrete, while the received signals corrupted by fading and noise approximately follow the Gaussian distribution. Inspired by this fact, an unsupervised clustering approach is proposed in \cite{GMMclustering} for the joint CE and MUD problem. Specifically, the Gaussian-mixture model (GMM) is used to model the noise-corrupted received signals, where each cluster in the GMM is associated with one data symbol. Then, the EM algorithm is adopted for this clustering problem. However, the successive interference cancellation (SIC) principle is adopted for MUD in \cite{GMMclustering}, which requires sufficiently large power difference among different users. For mMTC scenarios with densely deployed UEs, the received power of different UEs can be strongly correlated, which undermines the detection accuracy of SIC-based MUD. In addition, the variational Bayesian inference (VBI) method was employed for CE and MUD in one-bit quantized MIMO system \cite{VBIonebit}. With its powerful inference capability for intractable distributions, the VBI could effectively infer the distributions of the CSI and the data symbols from the received signals, which are heavily distorted after one-bit quantization.

\subsection{Motivations and Contributions}
Intrigued by the implementation feasibility, we consider the spreading technique to enable grant-free random access for mMTC in this paper. In order to address the deficiencies of existing solutions and improve the detection accuracy, an AMP-combined variational Bayesian inference-based clustering (AMP-VBIC) algorithm is proposed for joint user activity and data detection. Specifically, the decoupling operations in the AMP framework are adopted to mitigate multi-user interference (MUI) and decouple the signals of different UEs. Given the decoupled signals, we first \emph{extend} the data symbol alphabet to incorporate the null symbols from inactive UEs, and then formulate the joint user activity and data detection as a novel clustering problem under the GMM. Then, we develop a variational Bayesian inference based clustering (VBIC) algorithm for this clustering problem, where the CE result is also refined during the clustering procedure. The major contributions of this paper are summarized as follows.

(i) With the extended symbol alphabet, the joint user activity and data detection is formulated as a clustering problem under GMM. Then, we derive the VBIC algorithm for this clustering problem, which iteratively works in conjunction with the AMP decoupling module to refine the detection accuracy.

(ii) In the VBIC algorithm, the CE result is iteratively updated with the clustering result of all the data symbols, which in return improves the UAD and MUD accuracy. 

(iii) Analyses are provided to demonstrate the favorable linear complexity of the proposed AMP-VBIC algorithm, while simulation results show its superior detection accuracy over the state-of-art solutions. 

}

The remainder of this paper is organized as follows. Section \ref{System_section} describes the system model, and the AMP-VBIC algorithm is proposed in Section \ref{GFRAP} for the joint user activity and data detection problem. Simulation results are provided in Section \ref{simu_section}, and Section \ref{conclusion} concludes this paper.

\emph{Notations:} Scalar variables are written in italic letters. Vectors (or a set of variables) are written in boldface lower-case letters, and matrices in boldface upper-case letters. Unless stated otherwise, all the vectors are column vectors. $(\cdot)^T$ and $(\cdot)^H$ are the transpose and conjugate-transpose operations, respectively. $\bm{\text{E}}[\cdot]$ and $\bm{\text{Var}}[\cdot]$ take the expectation and variance of a random variable, respectively. $X\sim\mathcal{CN}(\mu,v)$ means that a random variable $X$ follows a complex Gaussian distribution with mean $\mu$ and variance $v$, and $\mathcal{CN}(x|\mu,v)$ is the probability density function (pdf) of this complex Gaussian distribution.  

\begin{figure}
	\centering
	\includegraphics[width=1\columnwidth]{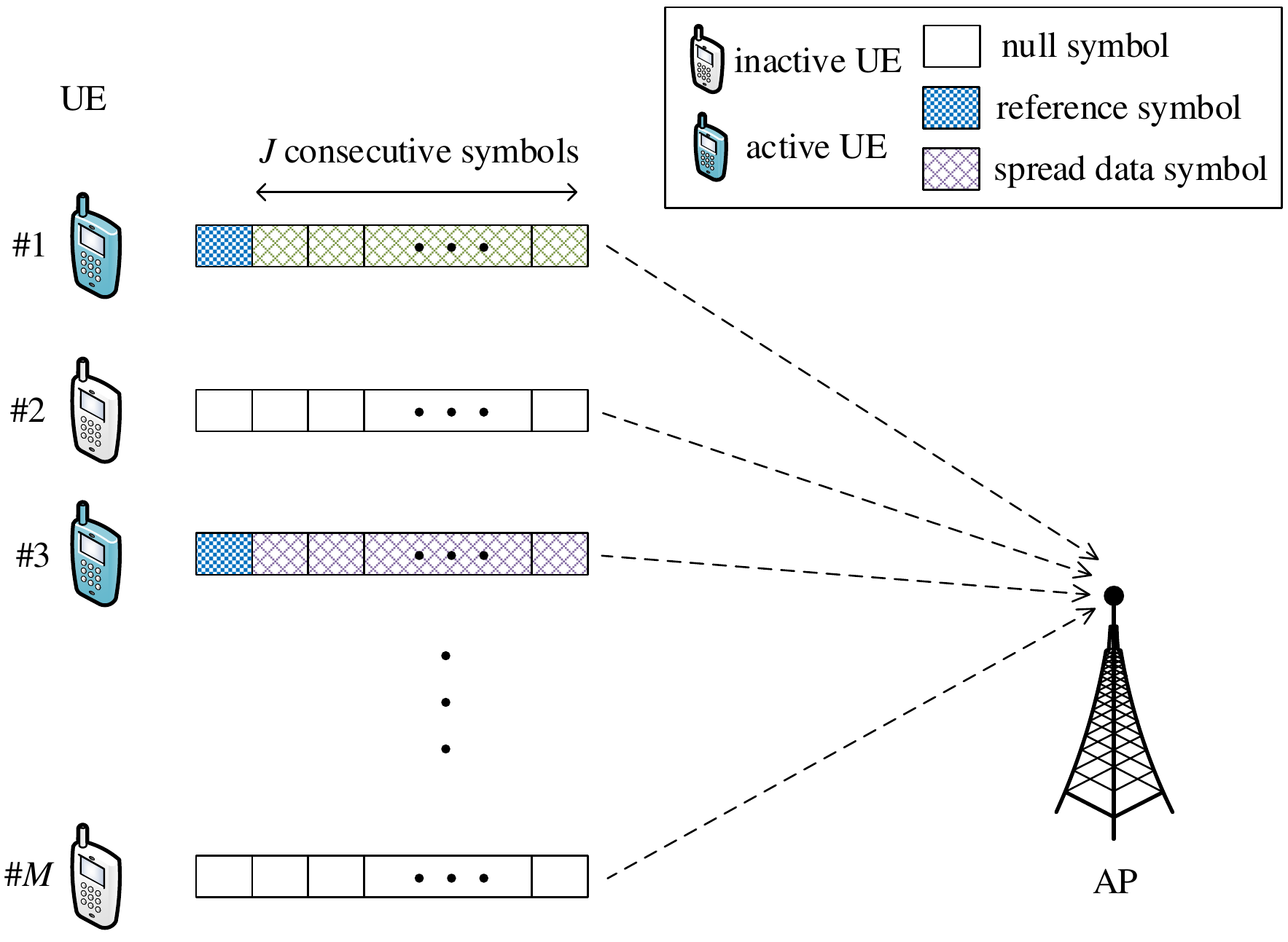}
	\caption{Grant-free random access system model.}
	\label{sys_mod}
\end{figure}
\section{System Model}\label{System_section}
As shown in Fig. \ref{sys_mod}, we consider a spreading-based uplink grant-free random access system with an access point (AP) serving $M$ user-equipments (UEs). Each UE is randomly activated with a probability $p_a$, while each active UE transmits $J$ consecutive symbols in one transmission block. The $j$-th modulated symbol of the $m$-th UE is denoted by $d^j_{[m]}$, which will be spread over a time-spreading sequence $\bm{a}_m$ of length $N$ before transmission. Then, the $j$-th received-signal vector $\bm{y}^j$ at the AP can be represented as 
\begin{equation}\label{rcvsig}
\bm{y}^j = \bm{A}\bm{U}\bm{d}^j  + \bm{w}^j,
\end{equation}
where $\bm{y}^j$ is the received signal vector with length $N$, and $\bm{A}=[\bm{a}_1,\bm{a}_2,\ldots,\bm{a}_M]$ is a $N\times M$ spreading matrix for all the $M$ UEs, $\bm{U}=\text{diag}\{\mu_1,\mu_2,\ldots,\mu_M\}$ is a diagonal matrix, and the $m$-th diagonal element $\mu_m\sim\mathcal{CN}(0,1)$ represents the Rayleigh channel coefficient of  the $m$-th UE. The term $\bm{d}^j=[d^j_{[1]},d^j_{[2]},\ldots,d^j_{[M]}]^T$ is the $j$-th transmitted symbol vector of $M$ UEs, and $\bm{w}^j$ represents the additive white Gaussian noise (AWGN) vector with length $N$. 

We assume an overloaded system with a large number of UEs, i.e. $N<M$. However, due to the sporadic activation of UEs in grant-free random access, there are only a small number of active UEs in each transmission frame. To facilitate the joint activity and data detection, we introduce an \emph{extended symbol alphabet} $\Delta=\{d_1, \Delta_a\}$ for the transmitted symbols $d^j_{[m]}$. Here, $d_1=0$ represents the equivalent \emph{null symbol} from inactive UEs, $\Delta_a$ is the modulation symbol alphabet of active UEs. For example, if Quadrature Phase Shift Keying (QPSK) modulation is adopted for transmission, we have $\Delta_a=\{\frac{1+1j}{\sqrt{2}},\frac{1-1j}{\sqrt{2}},\frac{-1+1j}{\sqrt{2}},\frac{-1-1j}{\sqrt{2}}\}$, where $j=\sqrt{-1}$. Furthermore, we denote $K$ as the size of $\Delta$, i.e. $\Delta=\{d_1,d_2,\ldots,d_{K}\}$.

Then, we consider the block transmission of $J$ consecutive symbols, and obtain a matrix version of (\ref{rcvsig}) as 
\begin{equation}\label{rcvmtx}
\bm{Y} = \bm{A}\bm{U}\bm{D}  + \bm{W}=\bm{A}\bm{X}  + \bm{W},
\end{equation}
where $\bm{Y}=[\bm{y}^1,\bm{y}^2,\ldots,\bm{y}^J]$ is the received signal matrix of size $N\times J$, $\bm{D}=[\bm{d}^1,\bm{d}^2,\ldots,\bm{d}^J]$ is the transmitted signal matrix of size $M\times J$, and $\bm{W}$ is an AWGN matrix of size $N\times J$. The spreading matrix $\bm{A}$ is known to the AP, and we assume a quasi-static block fading channel, i.e. the channel matrix $\bm{U}$ remains unchanged over the entire block of $J$ symbols. 

It is noted that $\bm{U}$ is unknown to the AP, and we define the \emph{intermediate detection target} $\bm{X}$ as $\bm{X}=\bm{U}\bm{D}$ in (\ref{rcvmtx}), from which the decision on $\bm{D}$ should be obtained. Since the data symbols in $\Delta_a$ are usually \emph{symmetric} for active UEs, we need to correct the \emph{phase ambiguity} when recovering $\bm{D}$ from $\bm{X}$. As shown in Fig. \ref{sys_mod}, we adopt a common solution to this phase ambiguity problem \cite{BSASP}, i.e. inserting a reference symbol (RS) before the data symbols. {\color{blue}More details on correcting this phase ambiguity problem will be later explained in Remark \ref{PRC} of Section \ref{GFRAP}.} In addition, each inactive UE equivalently transmits $J$ null symbols, i.e. $d^j_{[m]}=d_1=0$ for $j\in\{1,2,\ldots,J\}$. In this way, both $\bm{X}$ and $\bm{D}$ exhibit the \emph{row sparsity}. That is, the rows of $\bm{X}$ and $\bm{D}$ corresponding to inactive UEs only have zero elements, while the nonzero elements only reside in the rows corresponding to active UEs. The above-mentioned constraint is dubbed the \emph{joint sparsity} for the elements in $\bm{X}$ and $\bm{D}$, which will be used for activity detection. More details are explained as follows.

\section{Variational Bayesian Inference Clustering for Joint User Activity and Data Detection}\label{GFRAP}
To address the joint UAD and MUD problem, we derive the following AMP-VBIC algorithm. Typically, the operations in the AMP algorithm are divided into two modules, i.e. the \emph{decoupling module} which solves a linear mixing problem and a \emph{denoiser module} which usually functions as a demodulator for the data-detection target. However, it is shown in (\ref{rcvmtx}) that both the data matrix $\bm{D}$ and the unknown channel matrix $\bm{U}$ are included in the intermediate detection target $\bm{X}$. As a result, the demodulator in the typical AMP framework is not applicable to the detection of $\bm{X}$ under our model. As an alternative, we design the AMP-VBIC algorithm, where the denoiser module is now replaced with our proposed VBI clustering module. In this way, the VBI clustering module works in conjunction with the AMP decoupling module for the joint detection problem. The information exchange diagram between these two modules is illustrated in Fig. \ref{Info_ex}. More details are explained as follows.

\subsection{Pseudo Observation From AMP Decoupling Module}
For the linear mixing problem in (\ref{rcvmtx}) with known spreading matrix $\bm{A}$, the decoupling operations  of the AMP algorithm (i.e. the column-by-column operations in Algorithm \ref{alg:AMPVBI}) can be readily adopted to decouple the \emph{intermediate detection target} $\bm{X}$ from the received signal matrix $\bm{Y}$. That is, at the output of the AMP decoupling module, we can obtain a pseudo-observation matrix $\bm{R}$ for $\bm{X}$. Specifically, denote $r_m^j$ and $x_m^j$ as the element in the $m$-th row and $j$-th column of $\bm{R}$ and $\bm{X}$, respectively. The pseudo observation $r_m^j$ of the target element $x_m^j$ is written as
\begin{equation}\label{pseobs}
r_m^j = x_m^j  + n_m^j,
\end{equation} 
where the observation noise $n_m^j$ follows the distribution $n_m^j\sim\mathcal{CN}(0,\tau_m^j)$. Both $r_m^j$ and $\tau_m^j$ are provided by the AMP decoupling module. 

For each target element $x_m^j$, we have $x_m^j=\mu_md^j_{[m]}$. Since $\mu_m$ is unknown to the AP, the typical denoiser in the AMP algorithm (i.e. the demodulator) fails to demodulate $d^j_{[m]}$ from the observation $r_m^j$ of $x_m^j$. However, it is noted that each observation $r_m^j$ is associated with one specific data symbol $d_k\in\Delta$. Therefore, we can use the Gaussian mixture model, and cluster these observations by the following VBI-based clustering (VBIC) algorithm. After that, we update the mean $\hat{x}_m^j$ and variance $\hat{\tau}_m^j$ of $x_m^j$ in the VBI clustering module. All the mean $\hat{x}_m^j$ and variance $\hat{\tau}_m^j$ for $\forall m,j$ will compose a mean-value matrix $\bm{\hat{X}}$ and a variance matrix $\bm{\hat{T}}$, which will be fed into the AMP decoupling module for further refinement. 

\begin{figure}
	\centering
	\includegraphics[width=1\columnwidth]{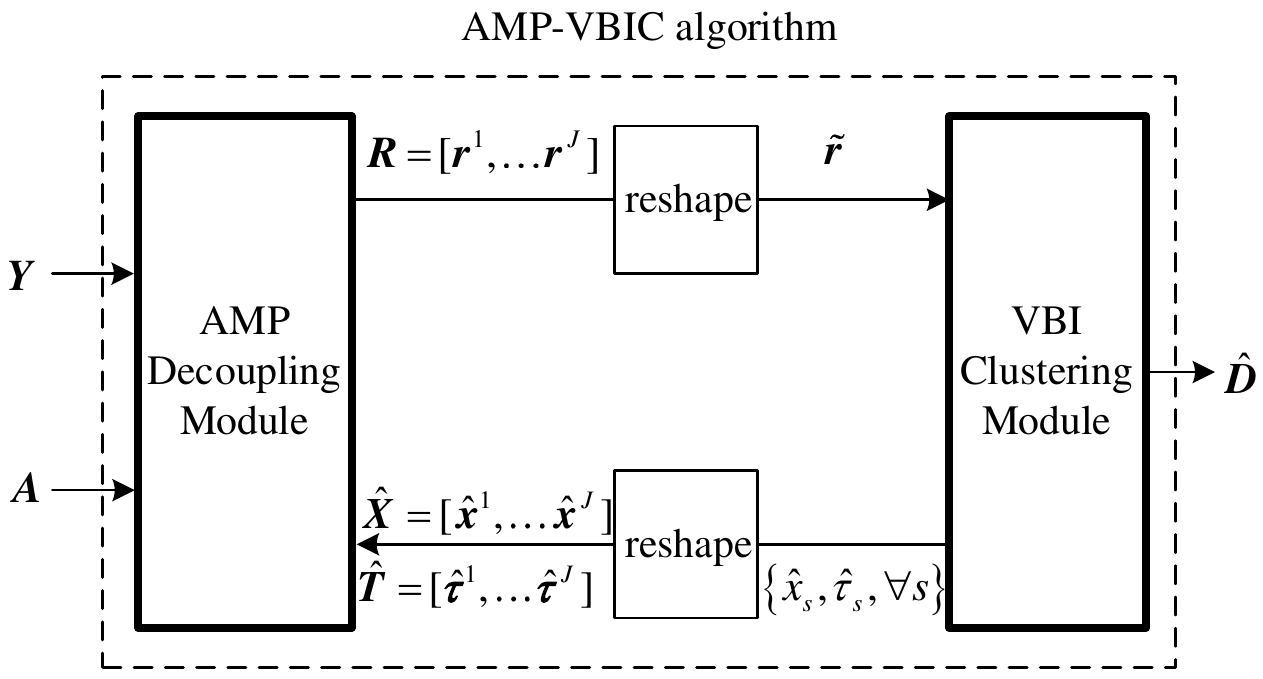}
	\caption{Information exchange between the AMP decoupling module and the VBI clustering module.}
	\label{Info_ex}
\end{figure}

\subsection{Data Detection in VBI Clustering Module}
For notational convenience, we first re-organize all the pseudo observations in $\bm{R}$ into a column vector $\bm{\tilde{r}}  =[\bm{r}^T_1,\bm{r}^T_2,\ldots,\bm{r}^T_M]^T$, where $\bm{r}^T_m$ is the $m$-th row vector of $\bm{R}$. Then, $\bm{\tilde{r}}$ is further denoted as $\bm{\tilde{r}}=[\tilde{r}_1,\tilde{r}_2,\ldots,\tilde{r}_S]^T$, where $S=MJ$ is the total number of observations. With a little abuse of notations, the observation $r^j_m$ in (\ref{pseobs}) is now re-written as $\tilde{r}_s$ in $\bm{\tilde{r}}$, where $s=j+(m-1)J$. In the following context, this relation among the observation index $s$, the symbol index $j$ and the UE index $m$ will always hold, unless stated otherwise. Then we assume that these pseudo observations $\tilde{r}_s$ are mutually independent with a Gaussian mixture model, i.e.
\begin{equation}\label{Gaumix}
{\color{blue}p(\bm{\tilde{r}}|\bm{\pi},\bm{\mu},\tau) = \prod\limits^S_{s=1}\Big(\sum\limits^K_{k=1}\pi_{sk}\mathcal{CN}(\tilde{r}_s|\mu_md_k,\tau^{-1})\Big)}
\end{equation}
where $d_k\in\Delta$, $\tau$ is a precision parameter, and $\pi_{sk}$ is the mixing coefficient. Here, the mixing coefficient $\pi_{sk}$ can be interpreted as the probability that the observation $\tilde{r}_s$ is associated with the data symbol $d_k\in\Delta$. We further denote $\bm{\mu}=\{\mu_1,\mu_2,\ldots,\mu_M\}$ as the collection of channel gains, and denote $\bm{\pi_s}=\{\pi_{s1},\pi_{s2},\ldots,\pi_{sK}\}$ as the collection of mixing coefficients for each observation $\tilde{r}_s$, while $\bm{\pi}=\{\bm{\pi_1},\bm{\pi_2},\ldots,\bm{\pi_S}\}$ denotes the collection of $\bm{\pi_s}$ for $\forall s$. For each observation $\tilde{r}_s$, we define a latent variable $\bm{z}_s$, which is a one-hot binary vector with length $K$. That is, $\bm{z}_s=[z_{s1},z_{s2},\ldots,z_{sK}]^T$, and only one element in $\bm{z}_s$ is 1. Here, $z_{sk}=1$ indicates the event that the observation $\tilde{r}_s$ is \emph{actually} associated with the symbol $d_k\in\Delta$. All the latent variables are collectively denoted as $\bm{Z}=[\bm{z}_1,\bm{z}_2,\ldots,\bm{z}_S]$, and we have the following conditional distributions
\begin{equation}\label{conddis}
p(\bm{\tilde{r}}|\bm{Z},\bm{\mu},\tau) = \prod\limits^S_{s=1}\prod\limits^K_{k=1}\mathcal{CN}(\tilde{r}_s|\mu_md_k,\tau^{-1})^{z_{sk}},
\end{equation}
and
\begin{equation}\label{conddis2}
p(\bm{Z}|\bm{\pi}) = \prod\limits^S_{s=1}\prod\limits^K_{k=1}(\pi_{sk})^{z_{sk}},
\end{equation}

To facilitate the following variational Bayesian inference, we first introduce the conjugate priors for the parameters $\bm{\mu}$, $\tau$, and $\bm{\pi}$. Specifically, we assume that the mixing coefficients $\bm{\pi}_s$ are mutually independent for different observation $\tilde{r}_s$, i.e. 
\begin{equation}\label{conj0}
p(\bm{\pi}) =\prod\limits^S_{s=1}p(\bm{\pi}_s).
\end{equation}
Then, we choose a Dirichlet prior distribution for $\bm{\pi}_s$, i.e.,
\begin{equation}\label{conj}
p(\bm{\pi}_s) = Dir(\bm{\pi}_s|\bm{\alpha}^s_0)={\color{blue}C(\bm{\alpha}^s_0)\prod\limits^K_{k=1}(\pi_{sk})^{\alpha_0^s[k]-1}},
\end{equation}
where $\bm{\alpha}^s_0=\Big[\alpha_0^s[1],\alpha_0^s[2],\ldots,\alpha_0^s[K]\Big]^T$ is the parameter vector, and $C(\bm{\alpha}^s_0)$ is a normalization constant for this Dirichlet distribution. By symmetry, we initialize the elements in $\bm{\alpha}^s_0$ by the same constant $\alpha_0$ for $\forall s$ and $\forall k$. Then, the conjugate priors for $\mu_m$ and $\tau$ are given by
\begin{equation}\label{conj2}
p(\bm{\mu}|\tau)=\prod\limits^M_{m=1}\mathcal{CN}\Big(\mu_m;\mu^m_0,(\lambda_0^m\tau)^{-1}\Big),
\end{equation}
\begin{equation}\label{conj3}
p(\tau)=Gam(\tau;a_0,b_0)=\frac{1}{\Gamma(a_0)}b_0^{a_0}\tau^{a_0-1}e^{-b_0\tau},
\end{equation}
where $\mu^m_0$ and $(\lambda_0^m\tau)^{-1}$ are the mean and variance for the complex Gaussian distribution of $\mu_m$, respectively. In addition, $a_0$ and $b_0$ are the parameters for the Gamma distribution of $\tau$, $\Gamma(\cdot)$ is the Gamma function. In this way, the joint distribution of all the variables are expressed as
\begin{equation}\label{joint_prob}
p(\bm{\tilde{r}},\bm{Z},\bm{\pi},\bm{\mu},\tau)=p(\bm{\tilde{r}}|\bm{Z},\bm{\mu},\tau)p(\bm{Z}|\bm{\pi})p(\bm{\pi})p(\bm{\mu}|\tau)p(\tau),
\end{equation}
Then, we consider the variational distribution $q$ of the latent variables and parameters with the following factorization
\begin{equation}\label{varidistr}
q(\bm{Z},\bm{\pi},\bm{\mu},\tau)=q(\bm{Z})q(\bm{\pi})q(\bm{\mu},\tau).
\end{equation}

Following the variational Bayesian inference procedure, we can infer different factor distributions in (\ref{varidistr}) as follows.

Firstly, for the latent variables $\bm{Z}$, we have
\begin{equation}\label{variZ}
\begin{split}
q(\bm{Z})\!&=\!\exp\!\Big({\color{blue}\bm{\text{E}}_{\bm{\pi},\bm{\mu},\tau}}\big[\ln p(\bm{\tilde{r}},\bm{Z},\bm{\pi},\bm{\mu},\tau)\big]+\text{const}\Big),\\
\!&=\!\exp\!\Big(\bm{\text{E}}_{\bm{\mu},\tau}\big[\ln p(\bm{\tilde{r}}|\bm{Z},\bm{\mu},\tau)\big]\!+\!\bm{\text{E}}_{\bm{\pi}}\big[\ln p(\bm{Z}|\bm{\pi})\big]\!+\!\text{const}\Big)\\
\!&=\!\prod\limits^S_{s=1}\prod\limits^K_{k=1}e^{z_{sk}}_{sk},
\end{split}
\end{equation}
where ``const" in (\ref{variZ}) refers to some constant numbers and they can be eliminated by the following normalization operations, $\bm{\text{E}}_{x}[f(x)]$ represents the expectation of $f(x)$ with respect to the random variable $x$, and the mixing coefficient $e_{sk}$ is normalized as follows for each observation index $s$, 
\begin{equation}\label{mixing_coeff_up} e_{sk}=\frac{\rho_{sk}}{{\sum\limits^K_{k^\prime=1}\rho_{sk^\prime}}},
\end{equation}
where
\begin{equation}\label{varirho}
\begin{split}
\rho_{sk}&\!=\!\exp\!\Big(\bm{\text{E}}_{\bm{\mu},\tau}\Big[\mathcal{CN}(\tilde{r}_s|\mu_md_k,\tau^{-1})\Big]+ \bm{\text{E}}_{\bm{\pi}}[\ln \pi_{sk}]\Big),\\
&\!\overset{(a)}{=}\!\!\exp\!\Big(\bm{\text{E}}_{\tau}[\ln \tau]\!-\!\ln\! \pi\!+ \!\bm{\text{E}}_{\bm{\pi}}[\ln \pi_{sk}]\!-\!\bm{\text{E}}_{\bm{\mu},\tau}\big[\tau|\tilde{r}_s\!-\!\mu_md_k|^2\big]\Big)\\
\end{split}
\end{equation}
where $\pi$ in equation ($a$) of (\ref{varirho}) is the circular constant, and the calculation of different terms in (\ref{varirho}) will be explained later as in (\ref{Episk})-(\ref{Emutau}).

After the derivation of $q(\bm{Z})$, we consider the distribution $q(\bm{\pi})$ in (\ref{varidistr}), and we have
\begin{equation}\label{varipi}
\begin{split}
q(\bm{\pi})&=\exp\Big(\bm{\text{E}}_{\bm{Z},\bm{\mu},\tau}\big[\ln p(\bm{Z}|\bm{\pi}) + \ln p(\bm{\pi})\big]+\text{const}\Big),\\
&\overset{(b)}{=}\exp\Big(\sum\limits^S_{s=1}\sum\limits_{k=1}^K\big(\bm{\text{E}}_{\bm{Z}}[z_{sk}]+\alpha_0^s[k]-1\big)\ln\pi_{sk}+\text{const}\Big),\\
&\overset{(c)}{=}\exp\Big(\sum\limits^S_{s=1}\sum\limits_{k=1}^K(e_{sk}+\alpha_0^s[k]-1)\ln\pi_{sk}+\text{const}\Big),\\
&\overset{(d)}{=}Dir(\bm{\pi}|\bm{\bar{\alpha}}),\\
\end{split}
\end{equation}
where we have $\bm{\text{E}}_{\bm{Z}}[z_{sk}]=e_{sk}$ in equation ($b$) of (\ref{varipi}), and $e_{sk}$ is calculated as in (\ref{varirho}). In addition, we can conclude from equation ($c$) of (\ref{varipi}) that $q(\bm{\pi})$ still takes the form of a Dirichlet distribution. With some manipulations on the constant terms, the updated Dirichlet distribution is given in equation ($d$), where the updated parameter vector $\bm{\bar{\alpha}}$ has components $\bar{\alpha}_0^s[k]$,
\begin{equation}\label{varialpha}
\bar{\alpha}_0^s[k]=\alpha_0^s[k]+e_{sk}.
\end{equation}

For the joint distribution $q(\bm{\mu},\tau)$ of $\bm{\mu}$ and $\tau$, we have 
\begin{equation}\label{varimutau}
\begin{split}
q(\bm{\mu},\tau)&\!\overset{(e)}{=}\!\exp\Big(\bm{\text{E}}_{\bm{Z},\bm{\pi}}\big[\ln p(\bm{\tilde{r}}|\bm{Z},\bm{\mu},\tau) \!+\! \ln p(\bm{\mu},\tau)\big]\!+\!\text{const}\Big),\\
&\!\overset{(f)}{=}\!\prod\limits^M_{m=1}\mathcal{CN}\Big(\mu_m;\bar{\mu}^m_0,(\bar{\lambda}_0^m\tau)^{-1}\Big)Gam(\tau;\bar{a}_0,\bar{b}_0).
\end{split}
\end{equation}

After some mathematical manipulations on equation ($e$) of (\ref{varimutau}), it is observed that $q(\bm{\mu},\tau)$ still takes the distribution form as in (\ref{conj2}) and (\ref{conj3}). The updated distribution parameters in equation ($f$) of (\ref{varimutau}) are calculated as
\begin{equation}\label{varilambda}
\bar{\lambda}^m_0=\lambda_0^m+\sum\limits^{mJ}_{s=(m-1)J+1}\sum\limits^{K}_{k=1}e_{sk}|d_k|^2,
\end{equation}
\begin{equation}\label{varimu}
\bar{\mu}^m_0=\Bigg(\lambda_0^m\mu_0^m+\sum\limits^{mJ}_{s=(m-1)J+1}\sum\limits^{K}_{k=1}e_{sk}d_k^*\tilde{r}_s\Bigg)\Big/{\bar{\lambda}^m_0},
\end{equation}
\begin{equation}\label{varia}
\bar{a}_0={a}_0+S,
\end{equation}
\begin{equation}\label{varib}
\bar{b}_0={b}_0+\sum\limits^M_{m=1}\lambda_0^m|\mu_0^m|^2+\sum\limits^S_{s=1}\sum\limits^K_{k=1}e_{sk}|\tilde{r}_s|^2-\sum\limits^M_{m=1}\bar{\lambda}_0^m|\bar{\mu}_0^m|^2,
\end{equation}
where $d_k^*$ is the conjugate of $d_k$, and the cumulative summation over $(m-1)J+1\leq s \leq mJ$ indicates that only the observations of UE $m$ are taken to update $\bar{\lambda}^m_0$ and $\bar{\mu}^m_0$. In addition, it is noted that $\bar{\mu}^m_0$ in (\ref{varimu}) represents the updated channel estimate for UE $m$. In other words, the data-detection result $e_{sk}$ is employed to refine the CE result $\bar{\mu}^m_0$. 

Now we can calculate different terms in equation ($a$) of (\ref{varirho})
\begin{equation}\label{Episk}
\bm{\text{E}}_{\bm{\pi}}[\ln \pi_{sk}]=\psi\Big(\bar{\alpha}_0^s[k]\Big)-\psi\Bigg(\sum\limits^K_{k^\prime=1}\bar{\alpha}_0^s[k^\prime]\Bigg),
\end{equation}
\begin{equation}\label{Etau}
\bm{\text{E}}_{\tau}[\ln \tau]=\psi(\bar{a}_0)-\ln\bar{b}_0,
\end{equation}
\begin{equation}\label{Emutau}
\begin{split}
&\bm{\text{E}}_{\bm{\mu},\tau}\big[\tau|\tilde{r}_s\!-\!\mu_md_k|^2\big]\\
&=\bm{\text{E}}_{\tau}\Bigg[\tau\bm{\text{E}}_{\bm{\mu}|\tau}\Big[|\tilde{r}_s|^2+|\mu_md_k|^2-2\text{Re}(\tilde{r}_s^*\mu_md_k)\Big]\Bigg],\\
&=\bm{\text{E}}_{\tau}\Big[\tau|\tilde{r}_s|^2\!+\!\tau|d_k|^2\Big(|\bar{\mu}^m_0|^2\!+\!(\bar{\lambda}^m_0\tau)^{-1}\Big)\!\!-\!2\tau\text{Re}(\tilde{r}^*_s\bar{\mu}^m_0d_k)\Big],\\
&=\frac{\bar{a}_0}{\bar{b}_0}\Bigg[|\tilde{r}_s|^2+|d_k|^2|\bar{\mu}^m_0|^2-2\text{Re}(\tilde{r}^*_s\bar{\mu}^m_0d_k)\Bigg]+\frac{|d_k|^2}{\bar{\lambda}^m_0},\\
\end{split}
\end{equation}
where (\ref{Episk}) and (\ref{Etau}) are obtained by the properties of the Dirichlet distribution and the Gamma distribution respectively, $\psi(\cdot)$ is the digamma function, $\tilde{r}^*_s$ is the conjugate of $\tilde{r}_s$, and $\text{Re}(\cdot)$ takes the real part of a complex number.

With (\ref{Episk})-(\ref{Emutau}), we can calculate $e_{sk}$ in (\ref{variZ}). After that, we update the mean $\hat{x}_s$ and variance $\hat{\tau}_s$ of each element  $x^j_m$ in the intermediate detection target $\bm{X}$. As illustrated in Fig. \ref{Info_ex}, the updated mean $\hat{x}_s$ and variance $\hat{\tau}_s$ will be fed back to the next iteration of AMP decoupling for further refinement. Specifically, the mean $\hat{x}_s$ of $x^j_m$ is updated as 
\begin{equation}\label{upmean}
\hat{x}_s=\bm{\text{E}}_{\bm{\mu},\tau,\bm{e_s}}[x^j_m]=\bm{\text{E}}_{\bm{\mu},\tau,\bm{e_s}}\Big[\mu_md^j_{[m]}\Big]=\bar{\mu}^m_0\sum\limits^K_{k=1}e_{sk}d_k,
\end{equation}
where $\bm{e_s}=\{e_{s1},\ldots,e_{sK}\}$, the random variables $\bm{\mu}$ and $\tau$ in (\ref{upmean}) takes the updated distribution as in (\ref{varimutau}), and $\bar{\mu}^m_0$ is the updated channel estimate for UE $m$, which is derived in (\ref{varimu}). Furthermore, the updated variance $\hat{\tau}_s$ of $x^j_m$ is derived as 
\begin{equation}\label{upvar}
\begin{split}
\hat{\tau}_s=&\bm{\text{Var}}_{\bm{\mu},\tau,\bm{e_s}}\big[x^j_m\big],\\
\overset{(g)}{=}&\bm{\text{Var}}_{\bm{\mu},\tau}\big[\mu_m\big]\bm{\text{Var}}_{\bm{e_s}}\big[d^j_{[m]}\big],\\
=&\bm{\text{E}}_\tau\Big[\bm{\text{Var}}_{\bm{\mu}|\tau}\big[\mu_m\big]\Big]\Big(\sum\limits^K_{k=1}e_{sk}|d_k|^2-\big|\sum\limits^K_{k=1}e_{sk}d_k\big|^2\Big)\\
=&\bm{\text{E}}_\tau\big[(\bar{\lambda}_0^m\tau)^{-1}\big]\Big(\sum\limits^K_{k=1}e_{sk}|d_k|^2-\big|\sum\limits^K_{k=1}e_{sk}d_k\big|^2\Big)\\
\overset{(h)}{=}&\frac{\bar{b}_0}{\bar{\lambda}_0^m(\bar{a}_0-1)}\Big(\sum\limits^K_{k=1}e_{sk}|d_k|^2-\big|\sum\limits^K_{k=1}e_{sk}d_k\big|^2\Big)\\
\end{split}
\end{equation}
where equation ($g$) of (\ref{upvar}) is obtained by the mutual independence of $\mu_m$ and $d^j_{[m]}$ in $x^j_m$, and equation ($h$) is obtained by the following property for a Gamma-distributed random variable $\tau\sim Gam(\tau;\bar{a}_0,\bar{b}_0)$, i.e.
\begin{equation}\label{Gamkmoment}
\bm{\text{E}}[\tau^k]=\frac{(\bar{b}_0)^{-k}\Gamma(\bar{a}_0+k)}{\Gamma(\bar{a}_0)}.
\end{equation}
\subsection{Exploiting Joint Sparsity for Activity Detection}
According to the VBI-based data-detection result, $e_{sk}$ in (\ref{variZ}) represents the probability that the observation $\tilde{r}_s$ belongs to the $k$-th cluster. We first ignore the joint sparsity, and the transmitted symbol $d^j_{[m]}$ should be decided as $d_k$ if $e_{sk}$ is the largest element among $\{e_{s1},e_{s2},\ldots,e_{sK}\}$. It is noted that $d_1=0$ is the equivalent null symbol from inactive UEs. We further denote $p^\text{act}_s$ and $p^\text{ina}_s$ as the probability that the symbol $d^j_{[m]}$ is transmitted from an active UE or an inactive UE, respectively. We have,
\begin{equation}\label{LLRs}
\begin{split}
p^\text{act}_s&\propto \text{max}\{e_{s2},\ldots,e_{sK}\},\\
p^\text{ina}_s&\propto e_{s1}.
\end{split}
\end{equation}

Denote $l^\text{VBI}_m$ as the VBI-based log-likelihood ratio (LLR) for the activity of UE $m$. Considering the joint sparsity caused by UE activity, $l^\text{VBI}_m$ is obtained from all the observations $\tilde{r}_s$ of UE $m$, i.e.
\begin{equation}\label{LLRm}
l^\text{VBI}_m=\!\!\sum\limits^{mJ}_{s=(m-1)J+1}\!\!\ln \frac{p^\text{act}_s}{p^\text{ina}_s}=\!\!\sum\limits^{mJ}_{s=(m-1)J+1}\!\!\ln\frac{\text{max}\{e_{s2},\ldots,e_{sK}\}}{e_{s1}}
\end{equation}

If $l^\text{VBI}_m$ is solely adopted for activity detection, the detection accuracy may be significantly undermined by the problem of \emph{false alarm}, which is explained as follows. 

For an inactive UE $m$, we have $d^j_{[m]}=0$, and therefore $x^j_m=\mu_md^j_{[m]}=0$. Consequently, the pseudo observation of $x^j_m$, i.e. $\tilde{r}_s$ will also be close to zero. In the VBI clustering module, $\tilde{r}_s$ is used to jointly estimate the unknown channel gain $\mu_m$ as in (\ref{varimu}) and update the mean $\hat{x}_s$ as in (\ref{upmean}). As a result, both the channel estimate result $\bar{\mu}^m_0$ and mean $\hat{x}_s$ will be close to zero. In this case, the VBI module may detect this inactive UE $m$ as an active UE which has a small channel gain $\bar{\mu}^m_0$. To address this problem, we consider an intuitive judgment that large estimate $\hat{x}_s$ usually comes from active UEs, while the VBI clustering module tends to produce small estimates $\hat{x}_s$ for inactive UEs. Then, according to the mean $\hat{x}_s$ and variance $\hat{\tau}_s$ in (\ref{upmean}) and (\ref{upvar}), we compute an \emph{offset} LLR \cite{offsetLEO,offsetEP} to improve the activity detection accuracy.

Specifically, we characterize the mean $\hat{x}_s$ as
\begin{equation}\label{meanrealerr}
\hat{x}_s=x^j_m+e^j_m,
\end{equation}
where $e^j_m$ denotes the estimation error between $\hat{x}_s$ and $x^j_m$, with the distribution $e^j_m\sim\mathcal{CN}(0,\hat{\tau}_s)$. For an inactive UE $m$, we have $x^j_m=0$, and therefore the \emph{prior} distribution of the mean $\hat{x}_s$ is $\hat{x}_s\sim\mathcal{CN}(0,\hat{\tau}_s)$. For an active UE $m$, the prior channel distribution $\mu_m\sim\mathcal{CN}(0,1)$ is assumed, and $\mu_m$ is independent from $d^j_{[m]}$. Therefore, the prior mean of $x^j_m$ is zero, while the prior variance of $x^j_m$ is calculated as
\begin{equation}\label{varxmj}
\bm{\text{Var}}(x^j_m)=\bm{\text{Var}}(\mu_{m})\bm{\text{Var}}(d^j_{[m]})=E_\text{sym}\overset{\Delta}{=}\frac{1}{K-1}\sum\limits_{k=2}^K|d_k|^2.
\end{equation}
In this way, if UE $m$ is active, the prior distribution for $\hat{x}_s$ is $\hat{x}_s\sim\mathcal{CN}(0,E_\text{sym}+\hat{\tau}_s)$. Based on the above-mentioned prior distribution of $\hat{x}_s$, we can calculate an offset LLR $l^\text{offset}_s$ for each observation index $s$
\begin{equation}\label{offsetLLR}
\begin{split}
l^\text{offset}_s&=\ln\frac{p(\text{mean$\ =\hat{x}_s$}|\text{UE $m$ is active})}{p(\text{mean$\ =\hat{x}_s$}|\text{UE $m$ is inactive})},\\
&=\ln\frac{\mathcal{CN}(\hat{x}_s;0,E_\text{sym}+\hat{\tau}_s)}{\mathcal{CN}(\hat{x}_s;0,\hat{\tau}_s)},\\
&=\ln\frac{\hat{\tau}_s}{E_\text{sym}+\hat{\tau}_s}+\frac{|\hat{x}_s|^2}{\hat{\tau}_s}-\frac{|\hat{x}_s|^2}{E_\text{sym}+\hat{\tau}_s}.
\end{split}
\end{equation}

Then, the decision LLR $l_m^\text{dec}$ for activity detection is obtained by combining the  VBI-based LLR $l_m^\text{VBI}$, the offset LLR $l^\text{offset}_s$, and the prior LLR $l_0=\ln\frac{p_a}{1-p_a}$ for each UE $m$,
\begin{equation}\label{decLLR}
l^\text{dec}_m=l_m^\text{VBI}+\sum\limits^{mJ}_{s=(m-1)J+1}l^\text{offset}_s+l_0.
\end{equation}

The data detection result is obtained as
\begin{equation}\label{DDdec}
\left\{
\begin{aligned}
&\hat{d}^j_{[m]}=d_{k^\prime}\ \text{where $k^\prime=\mathop{\arg\max}\limits_k\{e_{s2},\ldots,e_{sK}\}$, if $l^\text{dec}_m>0$}, \\
&\hat{d}^j_{[m]}=d_1=0, \text{if}\ l^\text{dec}_m\leq0.\\
\end{aligned}
\right.
\end{equation}

After traversing all the UE indexes $m$ and symbol indexes $j$ for $\hat{d}^j_{[m]}$, we finally obtain the detection result $\hat{\bm{D}}$ of the transmitted signal matrix $\bm{D}$. 

\subsection{Algorithm Summary and Complexity Analysis}
According to the explanations above, the AMP decoupling module works with the VBI clustering module to jointly detect UE activity and data symbols for the grant-free random access system. This entire framework is termed as the AMP-VBIC algorithm, and summarized as in Algorithm \ref{alg:AMPVBI}. 
\begin{algorithm}[t!]\setstretch{1.0}
	
	\caption{AMP-VBIC algorithm}
	
	\label{alg:AMPVBI}
	
	
	\KwIn{received signal matrix $\bm{Y}$, spreading matrix $\bm{A}$}
	
	
	\KwOut{data detection result $\hat{d}^j_{[m]} \ \text{for}\ \forall j,m$}
	
	{\textbf{Initialize:}} 
	
	{\ \ \ \ $\alpha_0=0.1,\ a_0=10^{-4},\ b_0=1,\ e_{sk}=\frac{1}{K}\ \text{for}\ \forall s,k$}
	
	{\ \ \ \ $\bm{S}=\mathbf{0}_{N\times J}$, $\bm{T_s}=\mathbf{0}_{N\times J}$, $\bm{P}=\mathbf{0}_{N\times J}$, $\bm{T_p}=\mathbf{0}_{N\times J}$.}
	
	{\ \ \ \ $\bm{R}=\mathbf{0}_{M\times J}$, $\bm{T}=\mathbf{0}_{M\times J}$, $\bm{\hat{X}}=\mathbf{0}_{M\times J}$, $\bm{\hat{T}}=E_\text{sym}\mathbf{1}_{M\times J}$.}
	
	\For{$l=1:N_{it}$}{
		
		\For{$j=1:J$}{
			$\bm{\tau^j_p}=|\bm{A}|^2\bm{\hat{\tau}}^j$
			
			$\bm{p^j}=\bm{A}\bm{\hat{x}^j}-\bm{\tau^j_p}\bm{\cdot}\bm{s^j}$
			
			$\bm{\tau^j_s}=\mathbf{1}\bm{./}(\bm{\tau^j_p}+\sigma_n^2\mathbf{1})$
			
			$\bm{s^j}=\bm{\tau^j_s}\bm{\cdot}(\bm{y^j}-\bm{p^j})$
			
			$\mathbf{1}\bm{./}\bm{\tau}^j=|\bm{A}^H|^2\bm{\tau^j_s}$
			
		    $\bm{r^j}=\bm{\hat{x}^j}+\bm{\tau}^j\bm{\cdot}(\bm{A}^H\bm{s^j})$
		}
		
		1. Reshape matrix $\bm{R}=[\bm{r^1},\ldots,\bm{r^J}]$ into vector $\bm{\tilde{r}}$.
		
		2. Update $\bar{\alpha}_0^s[k]$ as in (\ref{varialpha}) for $\forall s,k$
		
		3. Update $\bar{\lambda}^m_0$ as in (\ref{varilambda}) for $\forall m$
		
		4. Update $\bar{\mu}^m_0$ as in (\ref{varimu}) for $\forall m$
		
		5. Update $\bar{a}_0$ and $\bar{b}_0$ as in (\ref{varia}) and (\ref{varib})
		
		6. Update $e_{sk}=\rho_{sk}\big/\big({\sum\limits^K_{k^\prime=1}\rho_{sk^\prime}}\big)$ with $\rho_{sk}$ given in (\ref{varirho})
		
		7. $\alpha_0^s[k]=\bar{\alpha}_0^s[k], \lambda^m_0=\bar{\lambda}^m_0, \mu_0^m=\bar{\mu}^m_0, a_0=\bar{a}_0, b_0=\bar{b}_0$
		
		8. Update $\hat{x}_s$ in (\ref{upmean}) and reorganize $\hat{x}_s$ into matrix $\bm{\hat{X}}$.
		
		9. Update $\hat{\tau}_s$ in (\ref{upvar}) and reorganize $\hat{\tau}_s$ into matrix $\bm{\hat{T}}$.
		
	}
	\textbf{Data Detection:} Perform final data detection as in (\ref{DDdec})
	
	
\end{algorithm}

Specifically, $\alpha_0,a_0,b_0$ and $e_{sk}$ are initialized for the VBIC algorithm. The matrices $\bm{S}$, $\bm{T_s}$, $\bm{P}$, $\bm{T_p}$ and $\bm{T}$ are initialized and updated only within the AMP decoupling module, while their $j$-th columns are denoted as $\bm{s}^j$, $\bm{\tau_s}^j$, $\bm{p}^j$, $\bm{\tau}_p^j$ and $\bm{\tau}^j$, respectively. $\bm{R}=[\bm{r}^1,\ldots,\bm{r}^J]$ is updated in the AMP decoupling module, and then passed to the VBI clustering module. $N_{it}$ is the total iteration number, and we omit the iteration index $l$ in the notations for reading clarity. In addition, $|\bm{A}|^2$ returns the square of the modulus for each element in $\bm{A}$, while $\bm{\cdot}$ and $\bm{./}$ represent the element-wise multiplication and element-wise division operations, respectively. As shown in Algorithm \ref{alg:AMPVBI}, the $l$-th iteration of the AMP-VBIC algorithm starts with the decoupling module, i.e. an inner loop of column-by-column operations.  In this way, the AMP decoupling module accomplishes column-wise detection for all the $J$ columns in $\bm{X}$, and produces the pseudo observation matrix $\bm{R}=[\bm{r}^1,\ldots,\bm{r}^J]$. 

For the VBI clustering module, we first reshape the pseudo observation matrix $\bm{R}$ into vector $\bm{\tilde{r}}=[r_1,\ldots,r_S]^T$. Then the VBIC algorithm is performed as in line 2 to line 6, while line 7 initializes related parameters for the next VBIC iteration. Next, the mean $\hat{x}_s$ and variance $\hat{\tau}_s$ are updated in line 8 and line 9, and they will be reshaped into matrices $\bm{\hat{X}}=[\bm{\hat{x}}^1,\ldots,\bm{\hat{x}}^J]$ and $\bm{\hat{T}}=[\bm{\hat{\tau}}^1,\ldots,\bm{\hat{\tau}}^J]$, which will be fed back to the next AMP decoupling iteration. Finally, the data detection is made according to (\ref{DDdec}). 

We further analyze the computational complexity of the AMP-VBIC algorithm, which is dominated by the number of multiplication/division and exponential/logarithmic operations \cite{TwoStage,offsetLEO}. Firstly, the AMP decoupling operations are well-known for the low complexity. Considering all the $J$ columns, the update of $\bm{\tau^j_p}$, $\bm{p^j}$, $\bm{\tau}^j$, and $\bm{r^j}$ will introduce $\mathcal{O}(NMJ)$ multiplications, respectively. In addition, updating $\bm{\tau^j_s}$ and $\bm{s^j}$ entails only $\mathcal{O}(NJ)$ multiplications. For the VBIC algorithm, $\mathcal{O}(MJK)$ multiplications are required for the update of $\bar{\lambda}^m_0$, $\bar{\mu}^m_0$, $\bar{b}_0$, $\hat{x}_s$, and $\hat{\tau}_s$, respectively. Then, the calculation of $\rho_{sk}$ entails $\mathcal{O}(MJK)$ multiplications and $\mathcal{O}(MJK)$ exponential operations. Finally, it is concluded that the AMP-VBIC algorithm totally needs $\mathcal{O}(N_{it}NMJ+N_{it}MJK)$ multiplications and $\mathcal{O}(N_{it}MJK)$ exponential operations. In other words, the total complexity scales only linearly with the system parameters, making the AMP-VBIC algorithm computationally favorable for practical grant-free random access systems. 
{\color{blue}
\begin{remark}\label{PRC}
\textbf{(Phase Ambiguity and Correction by Reference Symbol)} The Gaussian mixture model in (\ref{Gaumix}) is employed in the VBIC algorithm, where the channel gain $\mu_m$ and data symbol $d_k$ are jointly estimated and detected. Since the modulation constellation is symmetric, we can always find non-zero phase shift $\theta$ satisfying $\hat{d}_k=d_ke^{j\theta}\in\Delta_a$, e.g. $\theta=\pi$. In this case, the VBIC algorithm may detect the data symbol as $\hat{d}_k=d_ke^{j\theta}$ and estimate the channel gain as $\hat{\mu}_m=\mu_me^{-j\theta}$ by mistake, since the wrong combination $(\hat{d}_k,\hat{\mu}_m)$ and the correct one $(d_k,\mu_m)$ will produce the same probability $p(\bm{\tilde{r}}|\bm{\pi},\bm{\mu},\tau)$ in (\ref{Gaumix}). This problem is dubbed as the phase ambiguity problem, and it can be readily addressed by the RS \cite{BSASP}. Specifically, we can take an arbitrary symbol from the modulation symbol alphabet $\Delta_a$ as the RS, and denote $d_{[m]}^\text{RS}$ as the RS of the $m$-th UE. Denote $\bm{d}^\mathcal{J}_{[m]}=\big[d^1_{[m]},\ldots,d^J_{[m]}\big]$ as the length-$J$ data sequence of the $m$-th UE, and $d_{[m]}^\text{RS}$ will be transmitted along with $\bm{d}^\mathcal{J}_{[m]}$ if the $m$-th UE is activated. If the $m$-th UE is further detected as active at the AP, the AMP-VBIC algorithm will produce the detection results $\hat{d}^\text{RS}_{[m]}$ and $\hat{\bm{d}}^\mathcal{J}_{[m]}$ for the RS and the data sequence, respectively. Since $d_{[m]}^\text{RS}$ is predetermined and known to the AP, the phase ambiguity can be corrected as
\begin{equation}\label{PRCequ}
\bar{\bm{d}}^\mathcal{J}_{[m]}=\hat{\bm{d}}^\mathcal{J}_{[m]}\frac{d^\text{RS}_{[m]}}{\hat{d}^\text{RS}_{[m]}},
\end{equation}
where $\bar{\bm{d}}^\mathcal{J}_{[m]}$ is the corrected data-detection result. For notation clarity, we assume that the final detection results obtained in (\ref{DDdec}) have already been corrected by the RS.
\end{remark}
}
\section{Simulations}\label{simu_section}
\begin{figure*}
	\centering
	\includegraphics[width=2\columnwidth]{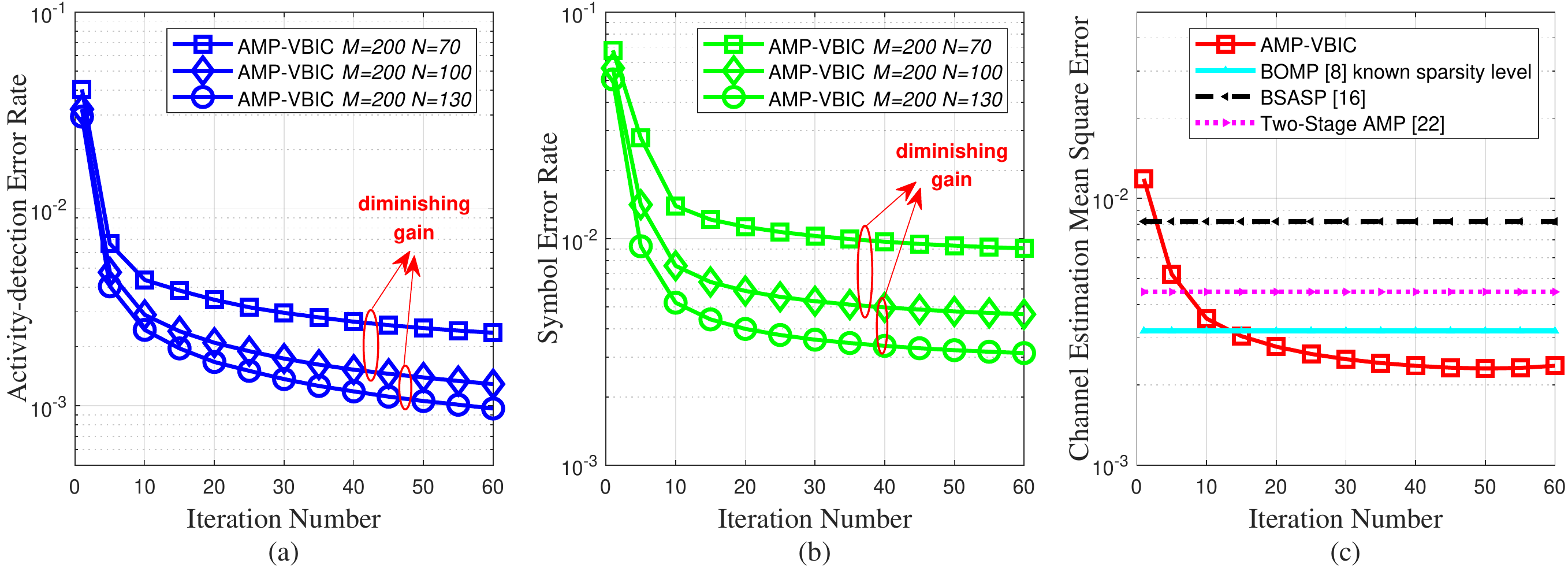}
	\caption{\color{blue}Joint detection performances under different iteration number $N_{it}$ with configurations $M=200$, $p_a=0.1$, $J=10$, and $\text{SNR}=5\text{dB}$.}
	\label{Convergence}
\end{figure*}
\begin{figure*}
	\centering
	\includegraphics[width=1.5\columnwidth]{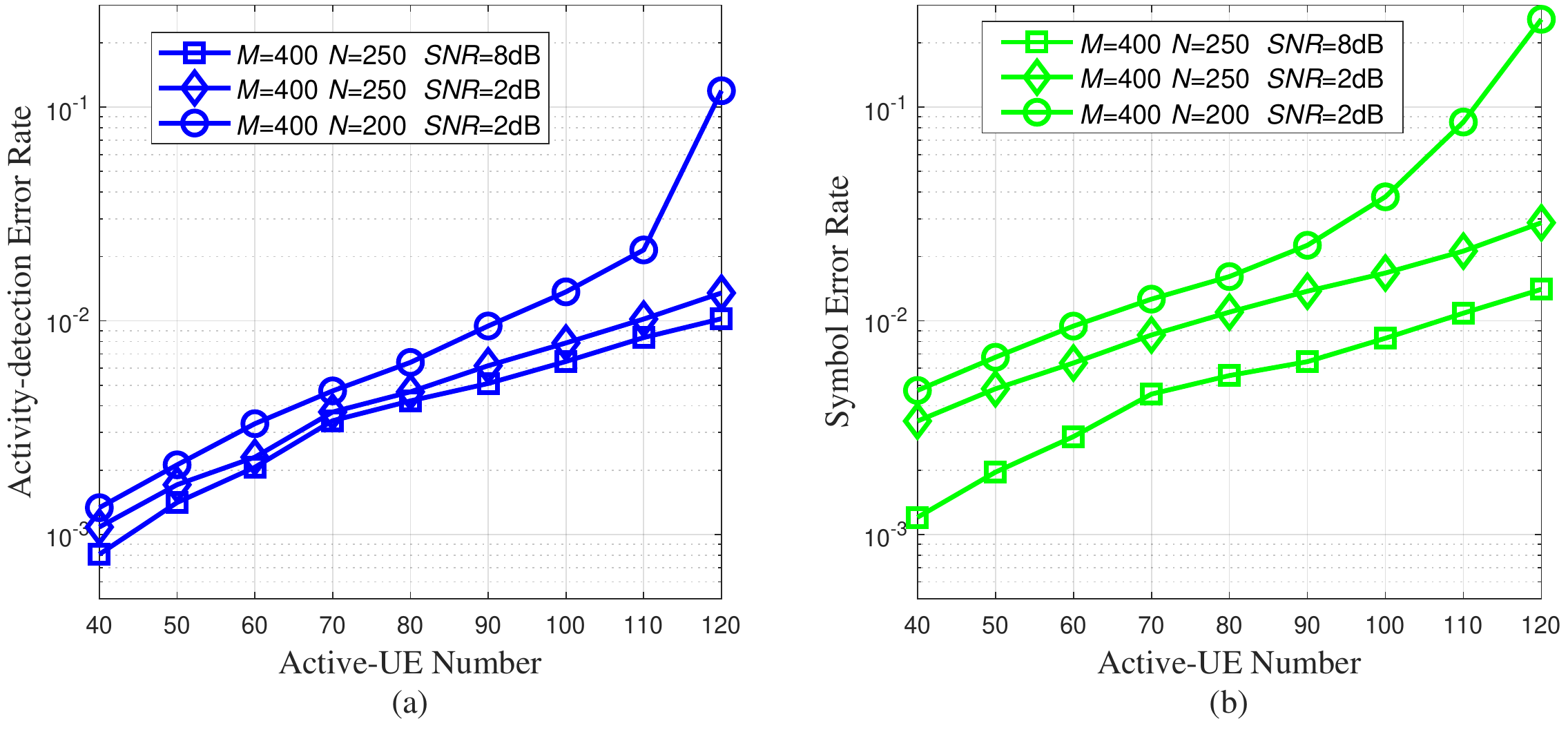}
	\caption{Joint detection performances under different number of active UEs with configurations $M=400$ and $J=10$.}
	\label{pa}
\end{figure*}

In this section, we evaluate the performances of our proposed AMP-VBIC algorithm for joint UE activity and data detection. {\color{blue}To begin with, we define the \emph{CSIR} as the knowledge of the channel matrix $\bm{U}$ at the AP, the \emph{support} of $\bm{X}$ as the \emph{exact identity} of active UEs, and the \emph{sparsity level} as the \emph{exact number} of active UEs. Due to the massiveness and sporadic activity of UEs, these three types of information defined above are unavailable to the AP. However, the spreading matrix $\bm{A}$ is assumed predetermined, and thus known to the AP. In addition, pseudo-random Gaussian sequences are adopted as spreading sequences for each UE, i.e. the elements in $\bm{A}$ are independently and identically distributed with distribution $\mathcal{CN}(0,1)$. Furthermore, we adopt the 16-Quadrature Amplitude Modulation (16-QAM) for transmitted data symbols, and we consider the detection performance for \emph{uncoded} data sequences\footnote{The proposed VBI-based data detection can also work with coded data sequences. If UE $m$ is detected as active, the probability that $d^j_{[m]}=d_k\in\Delta_a$ is proportional to $e_{sk}$ in line 6 of Algorithm \ref{alg:AMPVBI}. In this way, we can compute the LLR for each transmitted bit according to the 16-QAM constellation, and the LLR is output from the VBI module to the soft-decision decoder.}. Specifically, three performance metrics are considered in the following simulations, i.e., the activity-detection error rate (AER), the symbol error rate (SER), and the channel estimation mean square error (CE-MSE), which are defined as follows  
\begin{equation}\nonumber
\begin{split}
\text{AER}&\overset{\Delta}{=}{\sum\limits^M_{m=1}|\delta_m-\hat{\delta}_m|}/{M},\\
\text{SER}&\overset{\Delta}{=}\frac{1}{MJ}{||\bm{D}_{M\times J}-\hat{\bm{D}}_{M\times J}||_0},\\
\text{CE-MSE}&\overset{\Delta}{=}\sum\limits^M_{m=1}|\mu_m-\bar{\mu}_0^m|^2/M,
\end{split}
\end{equation}
where the activity indicator $\delta_m=1$ if the $m$-th UE is activated. Otherwise, $\delta_m=0$. $\hat{\delta}_m$ is the detection result of $\delta_m$. The $l_0$ norm of a matrix, i.e. $||\cdot||_0$ returns the number of non-zero elements, and we set $\mu_m=0$ for inactive UEs. In addition, we define the signal-to-noise ratio (SNR) as $\text{SNR}\overset{\Delta}{=}10\ln\frac{E_\text{Sym}}{\sigma_n^2}$.}
\begin{figure*}
\centering
\includegraphics[width=1.7\columnwidth]{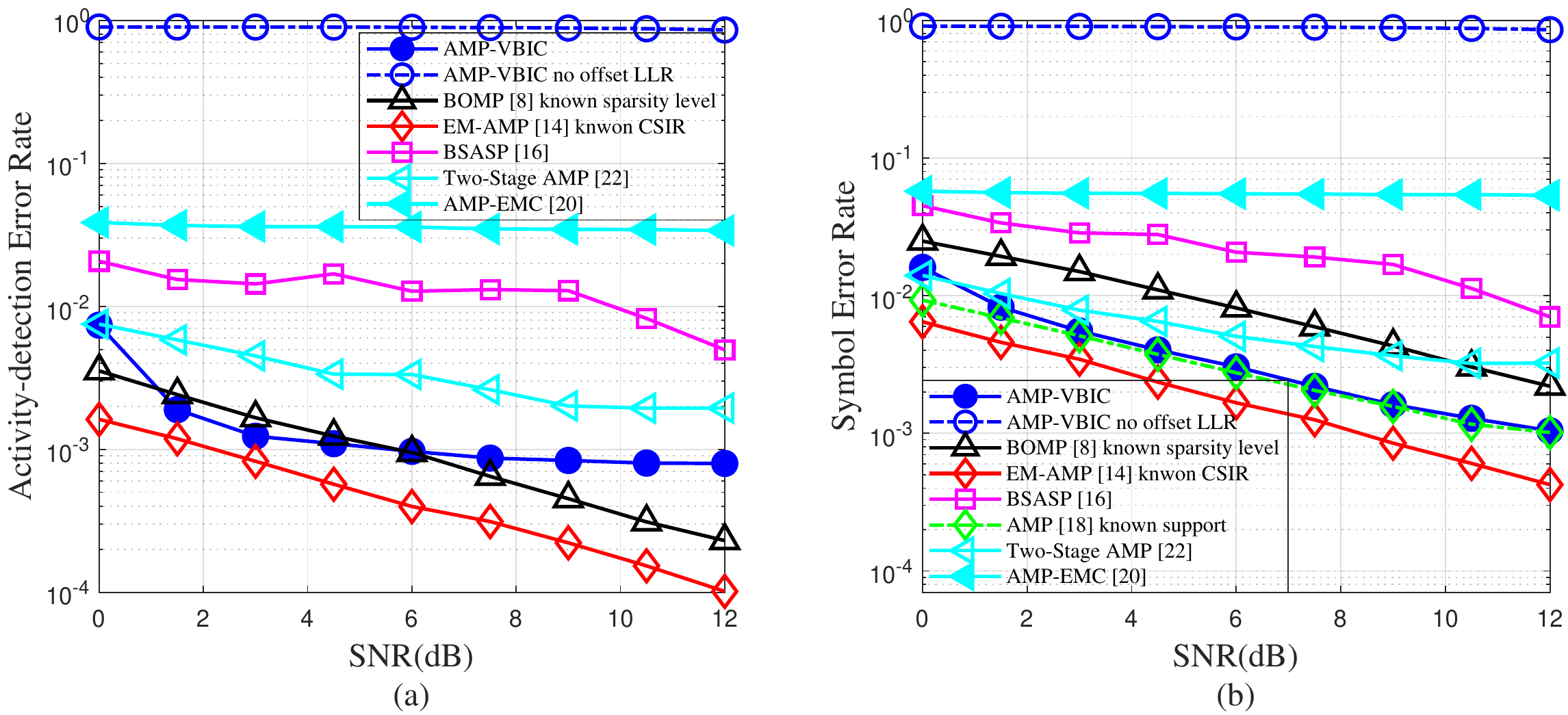}
\caption{\color{blue}Comparison on joint detection performances under different SNR with configurations $M=200$, $N=120$, $p_a=0.1$, and $J=20$.}
\label{SNR}\vspace{-0.1cm}
\end{figure*}
\subsection{Convergence Performance}
Firstly, we investigate the convergence performance of our proposed AMP-VBIC algorithm under different number of iterations $N_{it}$, and the simulation results are illustrated in Fig. \ref{Convergence}. It is shown in Fig. \ref{Convergence}(a) and Fig. \ref{Convergence}(b) that both the AER and the SER performances get rapidly improved in the first 20 iterations, then the detection performances tend to converge afterwards. Furthermore, increasing the spreading length $N$ could effectively lower the AER and SER. However, we also observe a \emph{diminishing gain}, i.e., increasing $N$ from 70 to 100 contributes to a more prominent performance gain than further increasing $N$ from 100 to 130. In addition, the spreading length $N$ is fixed as $70$ in Fig. \ref{Convergence}(c), and it is shown that we can gradually improve the CE accuracy with iterations. As shown in (\ref{varimu}), the CE update of $\bar{\mu}_0^m$ exploits the clustering results of \emph{all the data symbols}. Therefore, the AMP-VBIC algorithm outperforms the other existing solutions \cite{BOMP,BSASP,TwoStage}, which only employ one reference symbol for CE.

\subsection{Performance with Different Active-UE Number}
We further investigate the SER and AER performances of the AMP-VBIC algorithm with different number of active UEs, and the simulation results are illustrated in Fig. \ref{pa}. A general observation is that more active UEs will lead to deteriorated SER and AER performances. Then, we fix $N=250$ and reduce SNR from 8dB to 2dB. In this case, the performance loss on AER is almost negligible, while that on the SER performance is more obvious. After that, we fix SNR as 2dB, and reduce $N$ from 250 to 200. It is shown that, with only a small number of active UEs, the performance loss caused by reducing $N$ is not obvious on both AER and SER. However, given $N=200$ and $\text{SNR}=2\text{ dB}$, if we further increase the active-UE number from 90 to 120, both the SER and the AER performances will deteriorate drastically, which indicates the detection failure. In other words, to avoid detection failure caused by RA congestion, we need to increase the SNR or the spreading length $N$.

\subsection{Performance Comparison with Different SNR}
{\color{blue}Finally, we simulate the AER and SER performances under different SNR, and compare the proposed AMP-VBIC algorithm with different state-of-art solutions. These solutions include the BOMP algorithm \cite{BOMP} with \emph{known sparsity level}, EM-AMP algorithm \cite{AMPUADMUD} with \emph{known CSIR}, BSASP algorithm \cite{BSASP}, AMP algorithm \cite{AMP} with \emph{known support}, and the two-stage AMP algorithm where the UAD problem is firstly addressed by the BGMP algorithm \cite{TwoStage} and the MUD problem is secondly addressed by the AMP algorithm. In addition, we try to replace our proposed VBI clustering module with the EM-based clustering (EMC) approach in \cite{GMMclustering}, and establish an AMP-EMC algorithm for comparison. Recall that the sparsity level, the CSIR, and the support information are actually unavailable to the AP. Therefore, the solutions aided by such ideal knowledge can provide some performance lower bounds. The simulation results are illustrated in Fig. \ref{SNR}}

It is shown in Fig. \ref{SNR}(a) that the proposed AMP-VBIC algorithm exhibits superior AER performance to most solutions, except for those aided by known CSIR or sparsity level. In addition, the AER performance of the AMP-VBIC algorithm will not be further improved with higher SNR, which is consistent with the results in Fig. \ref{pa}(a). This observation can be explained by the fact that the channel noise has much smaller impacts on AER than the MUI in the high-SNR regime, while increasing $N$ is an effective method to mitigate the MUI. We can observe from Fig. \ref{SNR}(b) that the AMP-VBIC algorithm still outperforms most state-of-art solutions, and its SER performance could closely approach the performance lower bounds within a wide range of SNR. In contrast to the AER performance in Fig. \ref{SNR}(a), the SER of the AMP-VBIC algorithm could be effectively improved with higher SNR, since weaker channel noise is beneficial to the data-detection accuracy for active UEs. In addition, the BOMP algorithm \cite{BOMP} and the BSASP \cite{BSASP} algorithm are shown to exhibit inferior
SER performances, since they adopt the least square principle for data detection, which neglects the inherent modulation constraints of data symbols and thus undermines the data detection accuracy.

We also demonstrate the effectiveness of including the offset LLR for activity detection in (\ref{decLLR}). It is shown in Fig. \ref{SNR} that the AER and SER performances will approach $1-p_a$ if the offset LLR is not included for the AMP-VBIC algorithm. This result supports our claim that the problem of false alarm will significantly undermine the activity-detection accuracy if $l_m^\text{VBI}$ is solely adopted for activity detection. In addition, the simulation results demonstrate that EMC approach \cite{GMMclustering} fails to work for the clustering problem under the AMP framework. One possible reason is that the centroid of each cluster, i.e. the term $\mu_md_k$ in (\ref{Gaumix}), is estimated \emph{independently} in the EMC approach. In other words, the EMC approach neglects the inherent constraint that different cluster centroids of UE $m$ share the same channel-gain term $\mu_m$. Consequently, the clustering accuracy of the EMC approach is significantly undermined when the pseudo-observations $\bm{R}$ are contaminated by MUI, or when the symbol alphabet $\Delta$ is composed of high-order modulation symbols.

\section{Conclusions}\label{conclusion}
In order to address the joint UAD and MUD problem in grant-free random access, we formulated this joint detection problem as a clustering problem under the Gaussian mixture model. In conjunction with the AMP decoupling module, we developed a VBIC algorithm to solve this clustering problem. Compared with the state-of-art algorithms, our proposed AMP-VBIC algorithm demonstrated a significant performance gain.
\ifCLASSOPTIONcaptionsoff
  \newpage
\fi



%

\end{document}